\documentclass[conference]{IEEEtran}

\usepackage[english]{babel}
\usepackage[utf8]{inputenc}
\usepackage{amsmath}
\usepackage{multicol}
\usepackage{graphicx}
\usepackage{pbox}
\usepackage[T1]{fontenc}
\usepackage[utf8]{inputenc}
\usepackage[draft]{hyperref}
\usepackage{url}
\usepackage{authblk}
\usepackage{pbox}
\usepackage{epstopdf}
\usepackage{lmodern}
\usepackage[T1]{fontenc}
\usepackage{stfloats}
\usepackage{cite}
\usepackage{enumerate}
\usepackage{makecell}
\usepackage{diagbox}
\usepackage{algorithm}
\usepackage[bottom]{footmisc}
\usepackage[justification=centering]{caption}
\usepackage{paralist}
\usepackage[noend]{algpseudocode}
\usepackage[labelformat=simple, subrefformat=simple]{subcaption}

\title{\textbf{Edge-Fog Cloud: A Distributed Cloud for Internet of Things Computations}}

\author{
 Nitinder Mohan, Jussi Kangasharju\\Department of Computer Science, University of Helsinki, Finland\\Email: \{firstname.lastname@cs.helsinki.fi\}
}

\date{\today}
\makeatletter
\def\therule{\makebox[\algorithmicindent][l]{\hspace*{.5em}\vrule height .75\baselineskip depth .25\baselineskip}}%

\newtoks\therules% Contains rules
\therules={}% Start with empty token list
\def\appendto#1#2{\expandafter#1\expandafter{\the#1#2}}% Append to token list
\def\gobblefirst#1{% Remove (first) from token list
  #1\expandafter\expandafter\expandafter{\expandafter\@gobble\the#1}}%
\def\LState{\State\unskip\the\therules}% New line-state
\def\pushindent{\appendto\therules\therule}%
\def\popindent{\gobblefirst\therules}%
\def\printindent{\unskip\the\therules}%
\def\printandpush{\printindent\pushindent}%
\def\popandprint{\popindent\printindent}%

\algdef{SE}[WHILE]{While}{EndWhile}[1]
  {\printandpush\algorithmicwhile\ #1\ \algorithmicdo}
  {\popandprint\algorithmicend\ \algorithmicwhile}%
\algdef{SE}[FOR]{For}{EndFor}[1]
  {\printandpush\algorithmicfor\ #1\ \algorithmicdo}
  {\popandprint\algorithmicend\ \algorithmicfor}%
\algdef{S}[FOR]{ForAll}[1]
  {\printindent\algorithmicforall\ #1\ \algorithmicdo}%
\algdef{SE}[LOOP]{Loop}{EndLoop}
  {\printandpush\algorithmicloop}
  {\popandprint\algorithmicend\ \algorithmicloop}%
\algdef{SE}[REPEAT]{Repeat}{Until}
  {\printandpush\algorithmicrepeat}[1]
  {\popandprint\algorithmicuntil\ #1}%
\algdef{SE}[IF]{If}{EndIf}[1]
  {\printandpush\algorithmicif\ #1\ \algorithmicthen}
  {\popandprint\algorithmicend\ \algorithmicif}%
\algdef{C}[IF]{IF}{ElsIf}[1]
  {\popandprint\pushindent\algorithmicelse\ \algorithmicif\ #1\ \algorithmicthen}%
\algdef{Ce}[ELSE]{IF}{Else}{EndIf}
  {\popandprint\pushindent\algorithmicelse}%
\algdef{SE}[PROCEDURE]{Procedure}{EndProcedure}[2]
   {\printandpush\algorithmicprocedure\ \textproc{#1}\ifthenelse{\equal{#2}{}}{}{(#2)}}%
   {\popandprint\algorithmicend\ \algorithmicprocedure}%
\algdef{SE}[FUNCTION]{Function}{EndFunction}[2]
   {\printandpush\algorithmicfunction\ \textproc{#1}\ifthenelse{\equal{#2}{}}{}{(#2)}}%
   {\popandprint\algorithmicend\ \algorithmicfunction}%
\makeatother

\begin{document}
%-------IEEE Copy right form----------
\IEEEoverridecommandlockouts
\IEEEpubid{\makebox[\columnwidth]{978-1-5090-4960-8/16/\$31.00~
\copyright2016
IEEE \hfill} \hspace{\columnsep}\makebox[\columnwidth]{ }}
%----------------------------------------

\maketitle
\algnewcommand{\algorithmicgoto}{\textbf{Go to}}%
\algnewcommand{\Goto}[1]{\algorithmicgoto~\ref{#1}}%
\newcommand{\pushcode}[1][1]{\hskip\dimexpr#1\algorithmicindent\relax}

\begin{abstract}
Internet of Things typically involves a significant number of smart sensors sensing information from the environment and sharing it to a cloud service for processing. Various architectural abstractions, such as Fog and Edge computing, have been proposed to localize some of the processing near the sensors and away from the central cloud servers. In this paper, we propose Edge-Fog Cloud which distributes task processing on the participating cloud resources in the network. We develop the Least Processing Cost First (LPCF) method for assigning the processing tasks to nodes which provide the optimal processing time and near optimal networking costs. We evaluate LPCF in a variety of scenarios and demonstrate its effectiveness in finding the processing task assignments.
\end{abstract}


\section{Introduction}\label{sec:introduction}

%Cloud computing has created a radical shift in application computation and has evolved to provide low-cost and highly scalable computing services to its users. The cloud service providers deploy a network of large data centers spread across the globe. Applications with high computational requirements offload its processing tasks to a cloud service which first distributes the task to one of the servers and then provides the computed result to the application. However, due to their large sizes and operational requirements, the number of large data centers are quite less. For example, as of 2016, Amazon Web Services has only deployed 13 large data centers across the globe \cite{aws_loc}. Moreover, to further reduce their operational costs, the location of these data centers can skewed as the cloud service providers only deploy them at certain "fit" geographical locations. For example, Google has installed 7 of its 15 data centers in North America while there are only 2 data centers deployed in Asia and none in Africa \cite{google_loc}. More than often, the data center location is quite far from a "time-critical" client application and thus the benefit of reduced processing time in cloud server can be compromised by the network delay of offloading and retrieving the data from the cloud. 

Internet of Things (IoT) typically involves a large number of \emph{smart} sensors sensing information from the environment and sharing it to a cloud service for processing. A recent study by National Cable \& Telecommunications Association (NCTA) assumes that close to 50.1 billion IoT devices will be connected to the Internet by 2020~\cite{iot_study}. This leads to two major issues for computing IoT-generated data:
\begin{inparaenum}[i)]
\item the processing time of time-critical IoT applications can be limited by the network delay for offloading data to cloud, and
\item uploading data from a large number of IoT generators may induce network congestion thus incurring further network delay.
\end{inparaenum}

To tackle network issues involved in IoT and similar application's computation, researchers have proposed bringing the \emph{compute cloud} closer to data generators and consumers. One proposal is \textit{Fog computing cloud}~\cite{cisco_fog_overview} which lets network devices run cloud application logic on their native architecture. The objective of Fog cloud is to perform low-latency computation/aggregation on the data while routing it to the central cloud for heavy computations \cite{fogcomp_iot,mobile_fog}. On the other hand, \textit{Edge-centric computing cloud} \cite{edge_ccr} takes inspiration from projects such as SETI@Home, Folding@Home etc. \cite{Seti@Home,Folding@Home}, and proposes a consolidation of human-operated, voluntary resources such as desktop PCs, tablets, smart phones, nano data centers as a cloud. As the resources in Edge cloud usually lie in one-hop proximity to the IoT sensors; processing the data at the edge can significantly reduce the network delay ~\cite{edge_decentralized,edge_multimedia}.

While both Edge and Fog cloud envision bringing the cloud closer to the users, the two approaches only utilize their resources to carry out pre-processing tasks thus relying on a centralized cloud for heavy, computationally intensive tasks. This semi-dependence on a central cloud works well for applications which require tight data and compute coupling but proves disadvantageous for applications which generate large amounts of distributed data interactive user involvement.

In this paper, we present a node-oriented, fully decentralized hybrid of Edge and Fog compute cloud model, \textit{Edge-Fog cloud}. As the name suggests, the outermost layer of the Edge-Fog cloud is composed of a large number of volunteer, human-operated edge devices connected via ad-hoc network chains. The inner layer is composed of a dense network of Fog devices with high compute capabilities. Due to its decentralized architecture, the Edge-Fog cloud is capable of decoupling processing time from network delays by effectively handling processing close to the data generators. Edge-Fog cloud offers reliable data storage of raw and computed data at the central data store located at the core of its architecture.

The contributions we make in this paper are as follows:
\begin{itemize}
\item We present Edge-Fog cloud architecture, which is based on classifying compute devices into Edge and Fog layers, depending on their capabilities and ownership.
\item We design LPCF algorithm which assigns tasks on the available nodes in the Edge-Fog cloud while minimizing the processing time and network costs. We show that LPCF achieves near-optimal networking costs in polynomial time as opposed to exponential time complexity.
\item We develop an Edge-Fog cloud simulator and integrate it with LPCF assignment solver. We demonstrate and compare the efficiency of LPCF with its related works across a range of parameters and simulations.
\item We discuss and provide insights regarding the characteristics of Edge-Fog cloud  that will affect its performance in real-world.
\end{itemize}
%Our key contribution is to show how to assign tasks to the available nodes in the Edge-Fog cloud while minimizing the processing time and network costs. Our solution is based on the observation that by first minimizing the processing time, we can achieve near-optimal networking costs in polynomial as opposed to exponential time complexity. We demonstrate the efficiency of our solution across a range of parameters and discuss several deployment issues.

%We further propose efficient approach for assigning compute tasks on various available nodes of the edge-fog cloud. A processing task can be split up and deployed on devices with an associated cost. The goal of the approach is to assign the tasks in an efficient manner which also minimizes the overall cost of deployment. Furthermore, we evaluate and compare the efficiency of our task deployment approach and the edge-fog cloud to other similar approaches. 

The remainder of the paper is organized as follows. In Section~\ref{sec:edge-fog}, we describe the Edge-Fog cloud architecture. We propose our approach for deploying compute application tasks on the Edge-Fog cloud in efficient manner in Section~\ref{sec:deployment}. In Section~\ref{sec:evaluation} and~\ref{sec:discussion} we evaluate the effectiveness of our LPCF algorithm and discuss Edge-Fog deployment issues. We discuss the related work in Section~\ref{sec:relatedwork}. Section~\ref{sec:conclusion} concludes the paper.

\section{Edge-Fog Cloud}
\label{sec:edge-fog}

\subsection{Architecture}
\begin{figure}[!t]             
\centering
\captionsetup{justification=centering}
\includegraphics[width=0.32\textwidth]{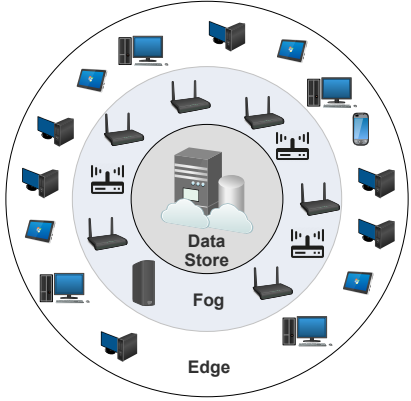}
\caption{\label{fig:arch}Proposed Edge-Fog cloud architecture}
\vskip -3mm
\end{figure}

Figure \ref{fig:arch} shows the architecture of the Edge-Fog cloud. Unlike the network-oriented view of the traditional cloud model, the Edge-Fog cloud takes a node-oriented approach wherein the model is divided into three layers comprising of different resource types. 

\textbf{Edge Layer.} The outermost layer of the cloud is Edge layer. The Edge is a collection of loosely coupled, voluntary\footnote{Several incentive/credit mechanisms can be employed for devices to volunteer as Edge resource. However, discussion of such mechanisms is currently out-of-scope of this paper.} and human-operated resources such as desktops, laptops, nano data centers, tablets, etc. As the name suggests, the resources reside at the edge of the network and are within one/two-hop distance from the IoT sensors and clients. Edge resources have varying ranges of computational capabilities from highly capable devices such as workstations, nano data centers etc. to less capable such as tablets or smart phones. Edge layer resources are assumed to have device-to-device connectivity within the layer and reliable connectivity to Fog layer. 

\textbf{Fog Layer.} The Fog layer resides on top of the edge and is a consolidation of networking devices such as routers and switches with high computing capabilities and ability to run cloud application logic on their native architecture. We envision Fog resources to be manufactured, managed and deployed by cloud vendors (such as CISCO\cite{cisco_fog}). As Fog layer forms the network backbone of Edge-Fog cloud, the resources in this layer are interconnected with high-speed, reliable links. Moreover, Fog resources reside farther from the edge of the network when compared to Edge layer but closer than a central cloud. Fog is used to effectively handle computationally intensive tasks offloaded by Edge resources.

\textbf{Data Store.} Unlike the traditional cloud model, the core of the Edge-Fog cloud has no computational capabilities and only serves as a repository for archiving all data in the cloud. A centralized store provides reliability and easy access to data by any computing resources in the cloud. Being at the core of the architecture, the Data Store is accessible by both Edge and Fog layers.

\subsection{Benefits of the Edge-Fog cloud}

The Edge-Fog cloud offer several benefits:

\begin{enumerate}
\item \textit{Reduced network load:} The Edge-Fog cloud provides computation at the edge of the network near the IoT generators thus reducing the amount of data that flows in the network. 
\item \textit{Native support for mobility:} Mobility along with reliability is a quintessential requirement for many IoT applications. Edge resources such as smartphones or laptops can offer native physical and virtual mobility for supporting such mobile IoT applications.
\item \textit{Providing context:} Resources in Edge-Fog cloud also provide contextual awareness to data generated by sensors. Edge resources play a role in combining data from sensors using location or application contexts. 
\item \textit{No single point of failure:} As computation in Edge-Fog cloud is completely decentralized, the model has no single point of failure. Several snapshots of an application can be deployed on the cloud for increased reliability. 
\end{enumerate}

Applications such as connected vehicles, energy monitoring, automated traffic control etc. can highly benefit from Edge-Fog cloud as most of the tasks in such applications are distributed and network-constrainted.

\section{Task Deployment on Edge-Fog Cloud}\label{sec:deployment}

\begin{figure*}[!t]
\centering
\begin{subfigure}{.28\textwidth}
  \centering
  \includegraphics[width=0.9\linewidth, height=0.5\linewidth]{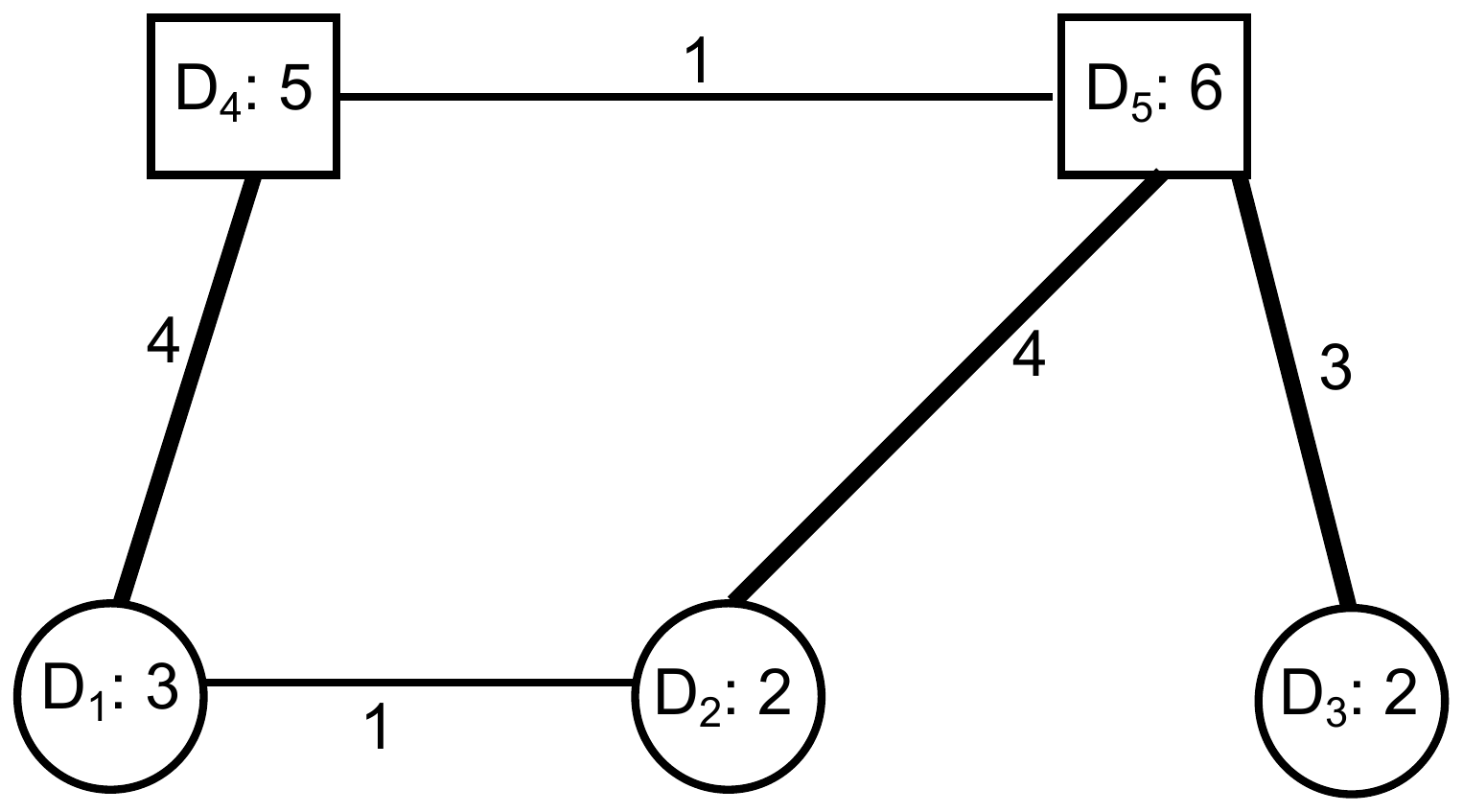}
  \caption{Edge-Fog cloud resource graph}
  \label{fig:EF_devices}
\end{subfigure}%
\hspace{0.1\textwidth}
\begin{subfigure}{.28\textwidth}
  \centering
  \includegraphics[width=0.9\linewidth, height=0.5\linewidth]{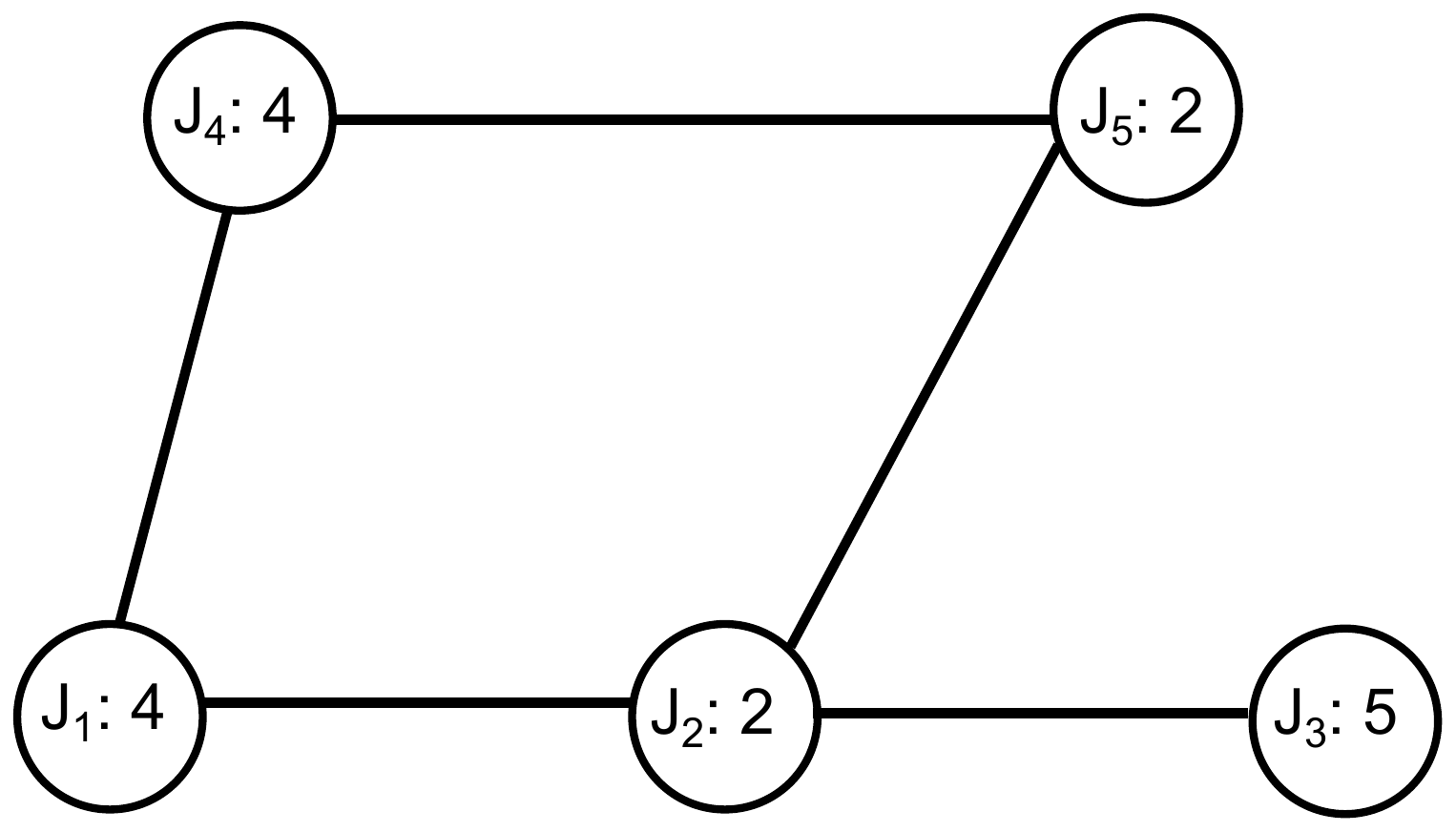}
  \caption{Available jobs dependence graph}
  \label{fig:EF_jobs}
\end{subfigure}
\caption{Deployment example. Each job in \autoref{fig:EF_jobs} needs to be deployed on a resource in \autoref{fig:EF_devices}.}
\label{fig:edge-fog_example}
\vskip -3mm
\end{figure*}

The Edge-Fog cloud is a scalable platform for a large number of interconnected Edge and Fog devices and efficiently utilize the processing power they offer. However, as the devices in the Edge-Fog cloud are governed by certain processing and network capabilities, deploying tasks on these devices has an associated cost. A typical task deployment algorithm must map a job node from the job graph to an Edge/Fog resource. The cost of deployment is dependent on both the properties of resources and that of the deployed task itself. For example, the more coordination needed by task with its peers for completion, the higher will be the associated network cost. In order to provide a scalable and efficient solution, the task deployment algorithm for Edge-Fog cloud should find the deployment snapshot with least possible cost without unduly impacting the overall completion time of that process.

Figure~\ref{fig:edge-fog_example} shows a snapshot of Edge-Fog cloud of three Edge and two Fog resources. Edge and Fog is represented by circular and rectangular nodes respectively. The link weight denotes the distance/communication cost between two devices. The processing power of each device is listed along with its label. Figure~\ref{fig:EF_jobs} shows the job graph to be deployed on the Edge-Fog cloud. We assume only two-way dependency between the jobs wherein Job \textit{$J_1$} and \textit{$J_2$} are dependent on each other if there exists a connection between them. The size of each job, listed along with its label, denotes the processing power required to complete the job. 

We assume that the number of tasks and devices to be equal while task deployment. If there are more tasks than devices, we split existing devices into virtual devices such that their number is equal to the number of tasks. In the opposite case, we simply ignore the superfluous devices.

\subsection{Network Only Cost (NOC) Assignment}

Previous works have tried to model task deployment algorithms which minimize the associated networking cost~\cite{haubenwaller_edge}. The formal definition of such task assignment strategies is to find an assignment which places \(\mathcal{N}\) jobs on \(\mathcal{N}\) devices such that the associated network cost is minimized. For the rest of the paper, we refer to all such algorithms as \textit{Network-Only Cost (NOC)} algorithms. Let \textit{$D_{conn}$(i,j)} represent the cost of connectivity between the devices \textit{$D_i$} and \textit{$D_j$} and \textit{$J_{conn}$(i,j)} denote the dependency between the jobs \textit{$J_i$} and \textit{$J_j$}. Both \textit{$D_{conn}$} and \textit{$J_{conn}$} are square matrices of size {$\mathcal{N}$x$\mathcal{N}$}. \textit{f(i)} signifies the constraint of assigning a particular job to a device.

With \(\mathcal{N}\) devices/jobs, the search space of possible assignments in NOC is \(\mathcal{N!}\). For example, in \autoref{fig:edge-fog_example}, the assignment \textit{$D_1\rightarrow J_1$; $D_2\rightarrow J_2$; $D_3\rightarrow J_3$; $D_4\rightarrow J_4$; $D_5\rightarrow J_5$} has network cost 17, whereas, the assignment \textit{$D_1\rightarrow J_4$; $D_2\rightarrow J_5$; $D_3\rightarrow J_3$; $D_4\rightarrow J_1$; $D_5\rightarrow J_2$} has cost 13. A naive NOC implementation would iteratively search for the assignment with least possible cost in the entire search space thus having the worst case complexity of O(\(\mathcal{N!}\)). On the other hand, NOC closely resembles the well-known Quadratic Assignment Problem (QAP) \cite{QAPLIB}. QAP generalizes minimal cost assignment as:

\begin{equation}
\sum_{i,j \in A} J_{conn}(i,j)*D_{conn}(f(i),f(j))
\label{QAP_eq}
\end{equation}
\qquad where \textit{A} is set of all arcs in the graph.

However, QAP is an NP-hard problem and its solution can only be approximated by applying constraints. Computing the optimal deployment for a problem space of 30 nodes using QAP may take up to a week on a computational grid comprising of 2500 machines~\cite{QAP_parallel}. Branch-and-bound based algorithms such as Gilmore-Lawler Bound (GLB) or Hungarian bounds can estimate the solution for small-sized QAP problems. Since the job scheduling on an Edge-Fog cloud may encompass computing an assignment of hundreds of devices, a more efficient algorithm for finding an optimal task assignment is needed.

\subsection{Least Processing Cost First (LPCF) Assignment}

As the Edge resources of the Edge-Fog cloud may not be highly processing-capable, the task assignment algorithm must also consider the associated processing cost of deployment. We thus propose LPCF, a task assignment solver which first minimizes processing cost of the assignment and further optimizes the network cost. In~\autoref{sec:evaluation} we show that LPCF algorithm is highly scalable when compared to NOC based algorithms. LPCF computes its optimal task assignment in the following manner.

\subsubsection{\textbf{Optimize the associated processing cost}}

LPCF calculates the processing cost associated with each possible assignment in the search space. The minimization function used by LPCF is:
\begin{equation}
\sum_{i,j \in A} C \bigg( \frac{J_{size}(i)}{D_{proc}(j)} \bigg) x_{ij}
\label{LAP_eq}
\end{equation}
\qquad where $C$ denotes the overall cost function; $J_{size}$ and $D_{proc}$ are matrices of size 1x\(\mathcal{N}\) representing the job sizes and the processing power of involved devices respectively. $x_{ij}$ is a binary job assignment variable.

Eq.~\ref{LAP_eq} is an objective function of Linear Assignment Problem (LAP) which unlike QAP, is polynomial~\cite{APsurveys}. Algorithms such as Kuhn-Munkres/Hungarian guarantee an optimal solution for this problem in $O(n^3)$ (worst case). The first step of LPCF employs such an algorithm to compute an assignment which has the least associated processing cost.  

\begin{table}[!t]
\centering
\begin{tabular}{|p{2cm}|c|c|c|c|c|c|c|c|c|}
\hline
Topology size & 5 & 10 & 15 & 30 & 60 & 100 & 150 \\
\hline
Original search space & 5! & 10! & 15! & 30! & 60! & 100! & 150! \\
\hline
LPCF search space & 1! & 3! & \textgreater4! & \textgreater5! & \textgreater7! & \textgreater8! & \textgreater9! \\
\hline
\end{tabular}
\caption{Problem search space reduction in LPCF}
\label{table:sub-problem-size}
\vskip -3mm
\end{table}

\subsubsection{\textbf{Reducing the sub-problem space size}}

As the Edge-Fog cloud consists of several homogeneous devices with similar processing capabilities, interchanging jobs assigned on any such two devices does not alter the associated processing cost. The same argument is also applicable  to homogeneous jobs in job graph. To illustrate, using equation \ref{LAP_eq} the assignment \textit{$D_1\rightarrow J_1$; $D_2\rightarrow J_2$; $D_3\rightarrow J_3$; $D_4\rightarrow J_4$; $D_5\rightarrow J_5$} in \autoref{fig:edge-fog_example} has the processing cost of 5.97 which remains the same if we interchange the jobs deployed on $D_1$ and $D_4$.

LPCF computes all possible compositions of the assignment computed in the first step and forms a smaller search space of assignments with least associated processing cost. \autoref{table:sub-problem-size} shows the reduction in problem search space achieved by LPCF. 

\subsubsection{\textbf{Accounting network cost of the assignment}} \label{subsec:compute}

In this step, LPCF computes the network cost associated with each assignment in the reduced problem search space and chooses the one with least network cost. Note that as the optimal assignment is updated at each iteration of the exhaustive search of sub-search space, a branch-and-bound variant of the algorithm can find the assignment within a time bound for large search space sizes. Thus, the assignment computed by LPCF has the least associated processing cost and \emph{almost} optimal network cost.  

Our approach has several advantages over NOC-based algorithms. The most fundamental of them is that unlike the NOC assignment, our algorithm guarantees an assignment in polynomial time thus significantly reducing the deployment calculation time. Moreover, as not all devices in the Edge-Fog cloud are highly processing capable, LPCF also takes into account the processing cost of the assignment.

\section{Evaluation}
\label{sec:evaluation}

We now evaluate the computation complexity for deploying jobs on several different Edge-Fog topologies. We have designed and implemented an Edge-Fog cloud simulator in Python (simulator code is available at \cite{EFSim}). The simulator generates a network of Edge and Fog resources and a job dependence graph based on several user-defined parameters. \autoref{table:values_simulator} shows the default parameter values we use for evaluating Edge-Fog cloud in this paper.  

\begin{table}[!t]
\centering
\begin{tabular}{|l|l|}
\hline
\rule{0pt}{3ex} 
\textbf{Properties}                                          & \textbf{Value}                     \\ \hline
\rule{0pt}{3ex}Total number of devices/jobs                     & \pbox{20cm}{Experiment \\specific}          \\
Number of Edge devices                              & 60\% of total             \\
Number of Fog devices                               & 40\% of total             \\
Processing power of Edge resources                & 2-5                       \\
Processing power of Fog resources                       & 7-9                       \\
Connection density in Edge layer (0-1)         & 0.2                       \\
Connection density in Fog layer (0-1)          & 0.6                       \\
\pbox{20cm}{Connection density between \\Edge and Fog layer (0-1)} & 0.5                       \\
Lowest job size in job pool                                 & 2                         \\
Highest job size in job pool                               & 6                         \\
Inter-dependence density between jobs (0-1)                 & 0.2                       \\ \hline
\end{tabular}
\caption{Default parameter values of Edge-Fog cloud simulator}
\label{table:values_simulator}
\vskip -3mm
\end{table}

We further implement and integrate LPCF task assignment solver in the Edge-Fog cloud simulator. To compare, we measure the performance of LPCF against two variants of NOC task assignment solver, permutation-based and QAP-based. For the QAP-based variant of NOC, we use an open-source implementation of Kuhn-Munkres solver available from QAPLIB~\cite{QAPLIB}.
\vskip -6mm
\subsection{Processing time analysis}

\begin{table*}[!t]
\centering
\begin{tabular}{|l|cccccccccc|}
\hline
\rule{0pt}{4ex} 
\pbox{20cm}{Number of \\Devices/Jobs =} & 5       & 10          & 15             & 20             & 30             & 40             & 50             & 60             & 100            & 150            \\ \hline
\rule{0pt}{3ex}NOC Permutation solver              & 0.068s  & 23m 20.785s & \textgreater1h & \textgreater1h & \textgreater1h & \textgreater1h & \textgreater1h & \textgreater1h & \textgreater1h & NA             \\
NOC QAP solver            & 0.026s  & 36.273s     & 3m 22.508s     & 18m 38.23s     & \textgreater1h & \textgreater1h & \textgreater1h & \textgreater1h & \textgreater1h & NA             \\
LPCF        & 0.0005s & 0.002s      & 0.044s         & 0.045s         & 0.18s          & 0.82s          & 4.358s         & 26.85s         & 7m3s           & \textgreater1h \\ \hline
\end{tabular}
\caption{Optimal assignment computation time}
\label{table:processing_time}
\vskip -3mm
\end{table*}

We analyze the overall processing time for computing an assignment by LPCF and NOC algorithms for several problem sizes. We set the maximum completion time of computation to one hour. The results are in \autoref{table:processing_time}. 

It is evident from the results that LPCF performs much better than both NOC-based solvers. For \textasciitilde30 node topology, where both solvers are unable to find an optimal assignment within the time limit, LPCF computes its assignment in under a second. For large topologies of \textasciitilde150 nodes, LPCF exceeds the maximum allotted time for the computing an optimal assignment. The primary reason for this increased computation time is due to the large size of the reduced search space size in LPCF. The current implementation of LPCF iteratively searches for the optimal assignment in reduced problem space which can be costly. However, a branch-and-bound variant of LPCF can significantly reduce the search time thus reducing the overall computation time.

\subsection{Comparative study of associated costs}

\begin{figure*}[t]
\centering
\begin{subfigure}{.28\textwidth}
  \centering
  \includegraphics[width=1\linewidth, height=0.7\linewidth]{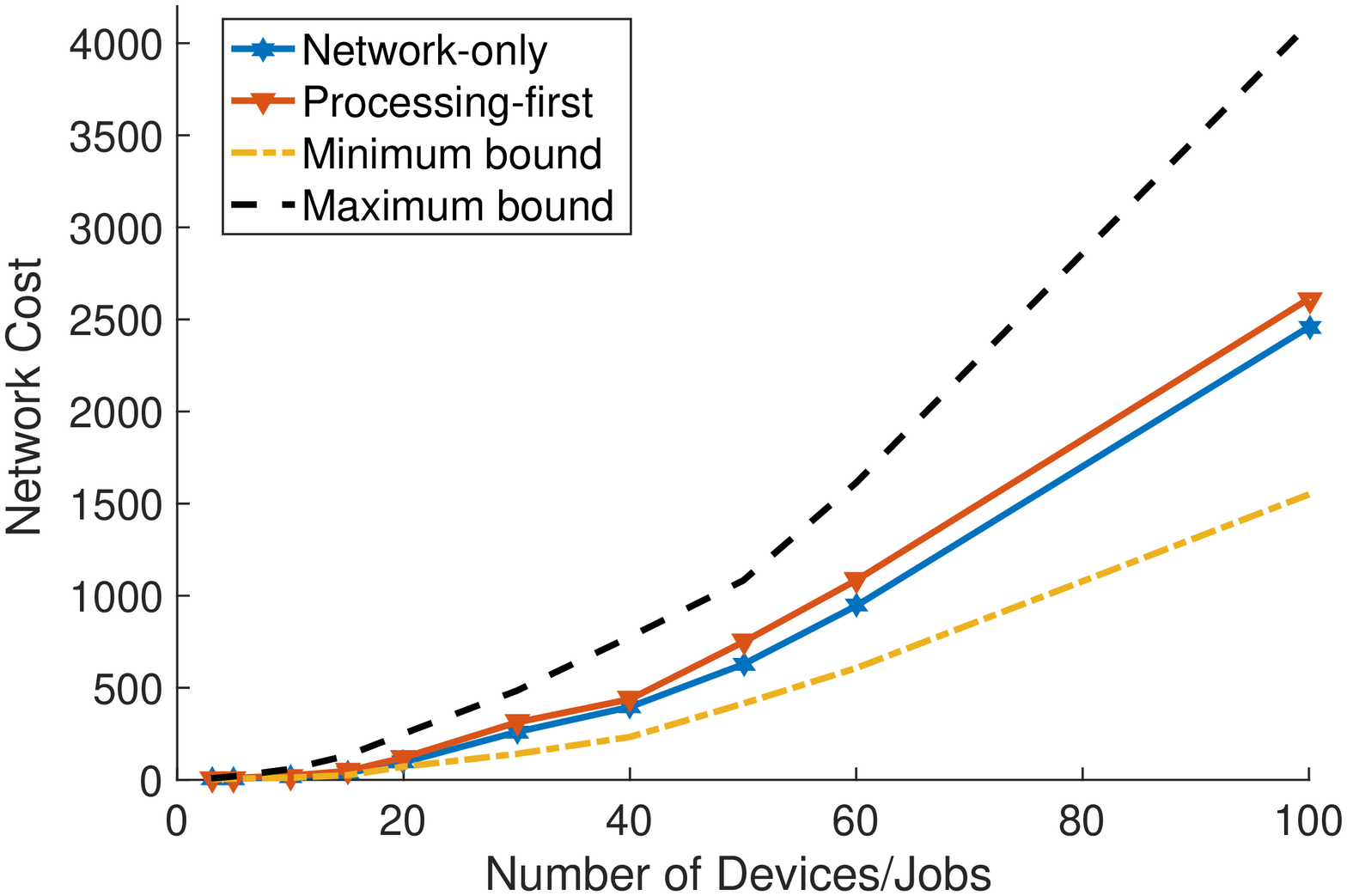}
  \caption{Network cost comparison between LPCF and NOC}
  \label{fig:netw_cost}
\end{subfigure}%
\begin{subfigure}{.28\textwidth}
  \centering
  \includegraphics[width=1\linewidth, height=0.7\linewidth]{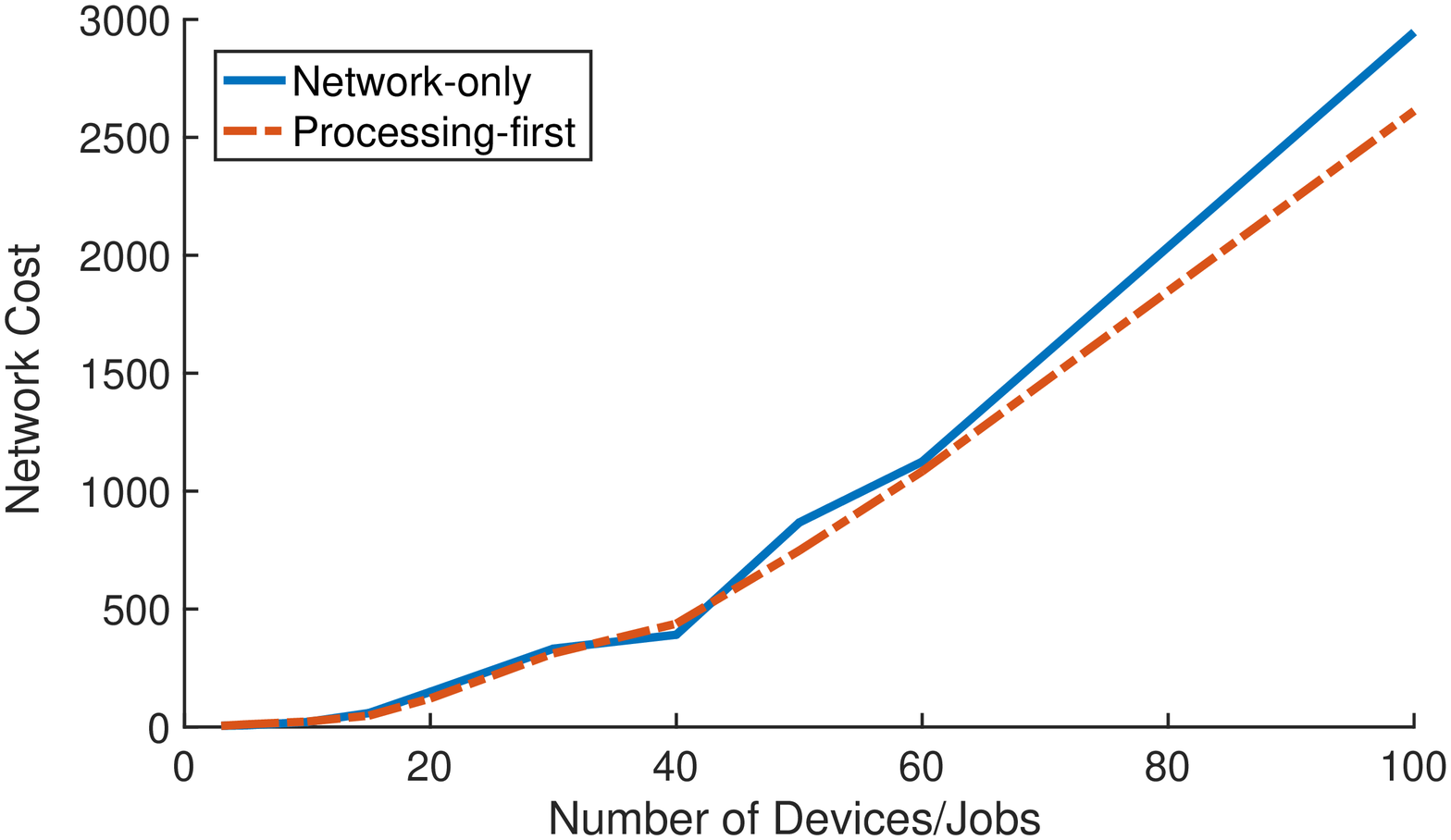}
  \caption{Network time comparison within time limit}
  \label{fig:bound_netw}
\end{subfigure}
\begin{subfigure}{.28\textwidth}
  \centering
  \includegraphics[width=1\linewidth, height=0.7\linewidth]{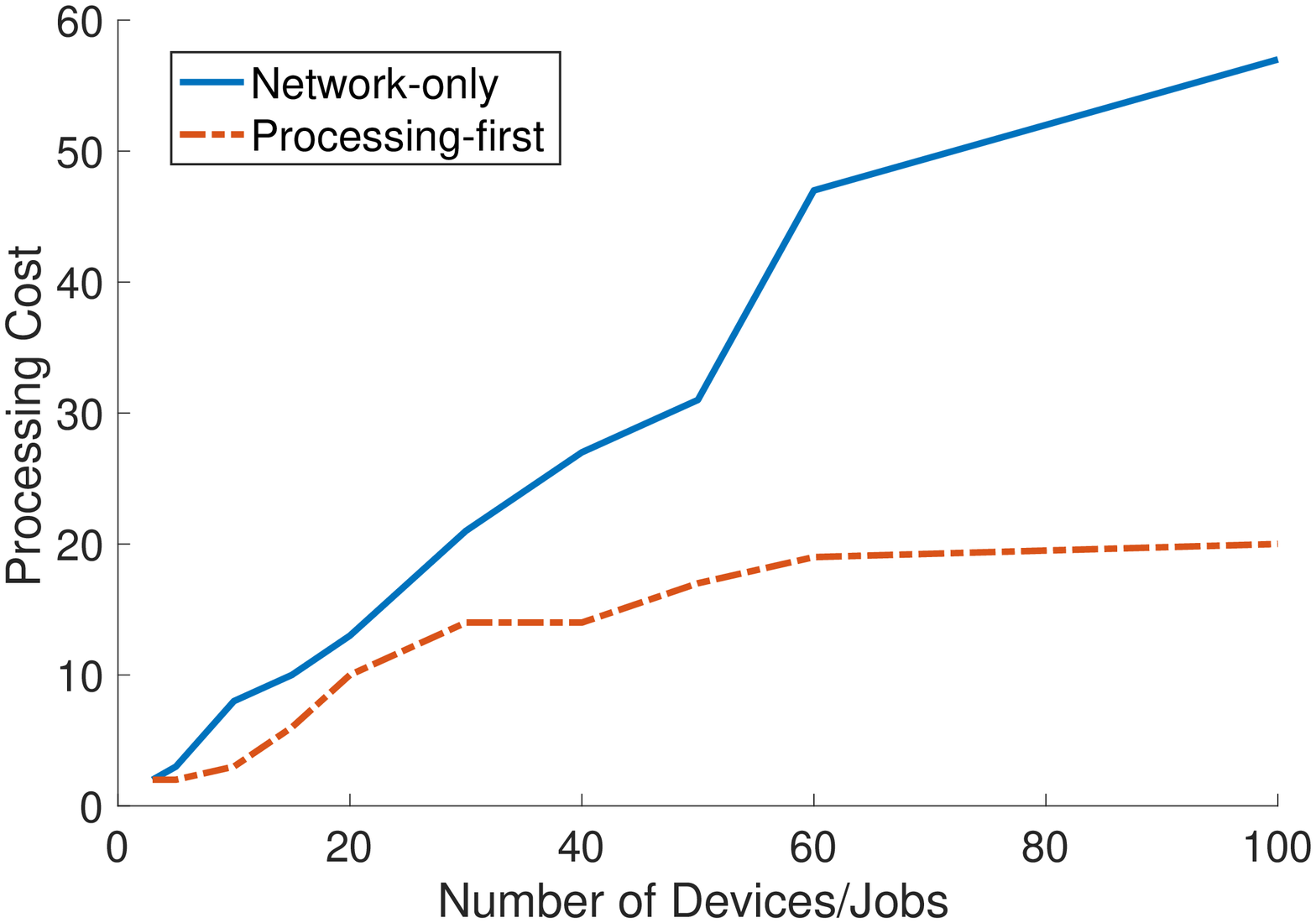}
  \caption{\label{fig:proc_cost}Processing cost analysis}
\end{subfigure}
\caption{Edge-Fog cloud associated cost analysis}
\label{fig:cost_analysis}
\vskip -3mm
\end{figure*}

\autoref{fig:netw_cost} compares the cost minimization achieved by LPCF when compared to NOC QAP task assignment solver. The minimum/maximum bounds are obtained by choosing the $\mathcal{N}$ smallest/largest link costs in the Edge-Fog cloud resource graph. It should be noted that the minimum/maximum cost depicted might not belong to a valid assignment as it does not consider job dependencies.

It is evident from the figure that even though the assignment computed via LPCF first optimizes processing time for assignment, the associated network cost is within 10\% range of the optimal value computed by the NOC. Also, we can see from \autoref{table:values_simulator} that the QAP-based NOC solver have significantly higher computing time when compared to LPCF. We further implemented a branch-and-bound variant of QAP solver which approximates the best solution within the specified time limit. We then limit the computation time of QAP to that of LPCF and plot the associated network costs of the optimal assignments found by both these algorithms. The plot is shown in \autoref{fig:bound_netw}. Here we see that for large topologies, the assignment computed by LPCF has lower associated network cost than that computed by NOC.

Figure \ref{fig:proc_cost} compares the associated processing cost of assignments computed by the two solvers. As unlike NOC, the assignment computed via LPCF is optimized on processing cost, the associated processing cost of the assignment computed by LPCF is always lower than that computed by NOC.

\section{Discussion} \label{sec:discussion}

\textbf{Q1. Which node is responsible for running the assignment solver?}

The LPCF algorithm needs a centralized controller for managing the execution of the algorithm, however, its actual execution can be distributed and is not dependent on any single node. One node needs to be able to get the snapshot of the system state (availability of nodes and costs of links) and we assume this snapshot to remain constant during the execution of the algorithm. Calculating the individual assignment permutations for processing or networking costs in steps 1 and 3 can be distributed to other nodes or can be performed by the controller. The LPCF algorithm can thus be executed by any of the nodes in the system, whether an Edge node or a Fog node. We do not consider the cost of running the algorithm in our evaluation since the overheads are similar for both LPCF and NOC QAP (namely obtaining the snapshot and going through the permutations).

\begin{figure}[!t]             
\centering
\captionsetup{justification=centering}
\includegraphics[width=0.28\textwidth]{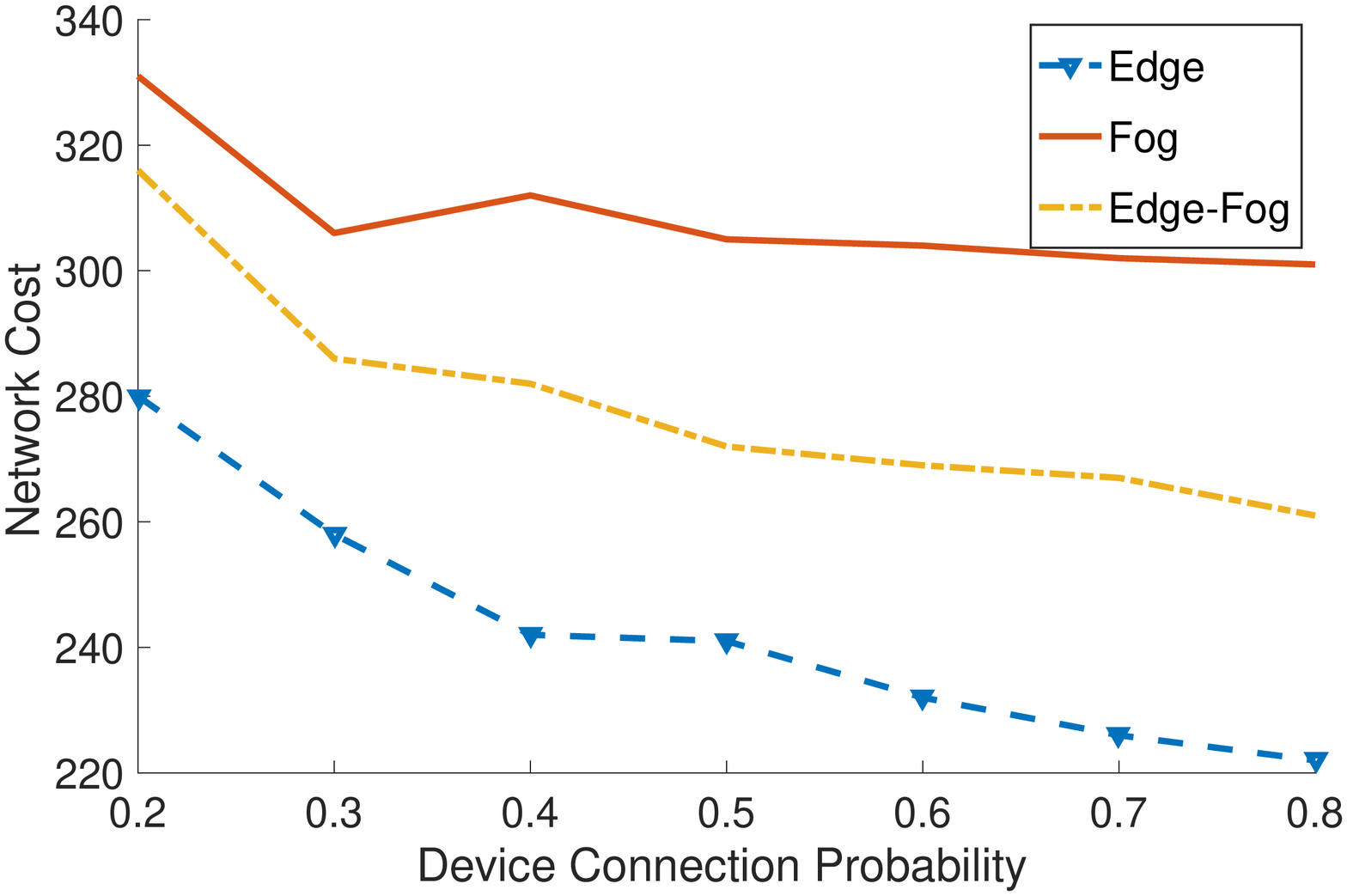}
\caption{\label{fig:netw_device}Network cost variation with inter-device connection densities}
\vskip -5mm
\end{figure}

\textbf{Q2. How well should the devices in the Edge-Fog cloud be connected to each other?}

Edge layer has optimistic connections within itself whereas the Fog layer has dense network connections; but the inter-layer connections between the Edge and the Fog are much higher cost and spans multiple hops. We try to find the optimal connection density of each layer (and inter-layer connections as well) such that the resulting assignment has low associated network cost. We increase the connection density of each layer from 20\% to 80\% and plot the changes in network cost of assignment computed by LPCF in Figure~\ref{fig:netw_device}.

As we increase the connection density of Edge, Fog and interconnections we see a decrease of \textasciitilde21\%, \textasciitilde9\% and \textasciitilde17\% in network cost respectively. The inter-layer connections play a major part in resulting network cost; increasing their density impacts the overall cost much more. We can infer that deploying jobs in an Edge-Fog cloud which have well-connected devices in edge layer and dense connections between edge and fog layers, the overall cost of deployment is significantly reduced.

\begin{figure}[!t]             
\centering
\captionsetup{justification=centering}
\includegraphics[width=0.28\textwidth]{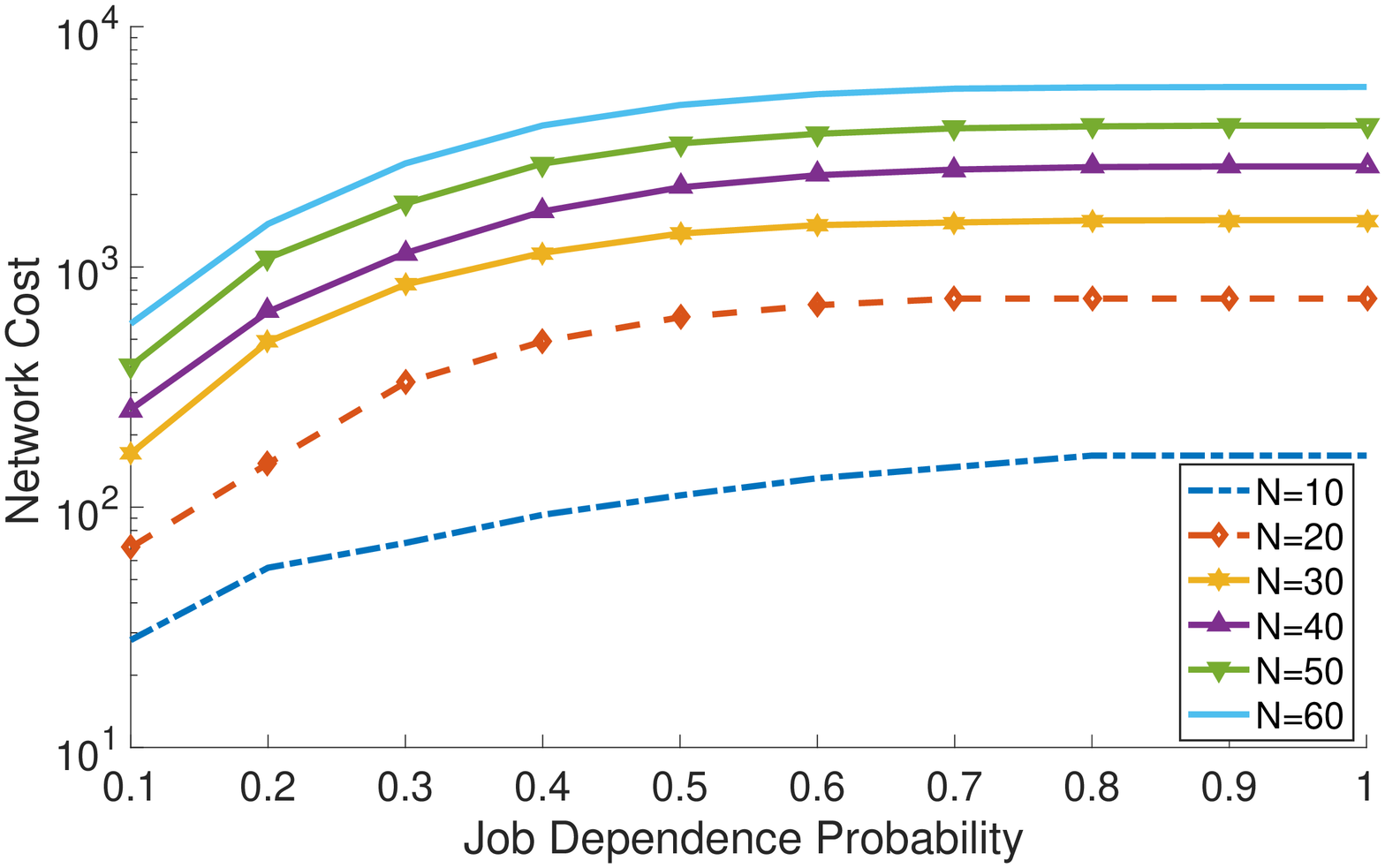}
\caption{\label{fig:netw_job}Effect of job dependence on network cost of assignment}
\vskip -5mm
\end{figure}

\textbf{Q3. Do the properties of job graph deployed on the Edge-Fog cloud also affect the overall cost?}

In Figure~\ref{fig:netw_job}, we change the interdependence of the job graph deployed on the Edge-Fog cloud from 10\% to 100\% and calculate the network cost associated with the deployment. We then deploy the job graph on several topology sizes of Edge-Fog cloud.

The results clearly show that higher dependence between the jobs result in a higher network cost. This is because the dependence links between the sub-jobs are mapped to the links between the devices of the Edge-Fog. Larger dependence links map to a mesh of device linkages thus leading to an increased network cost. It can also be seen from the figure that after a particular job dependency value, the associated network cost of assignment stabilizes. This is primarily because after a particular job inter-dependence all heavy links of the device graph are part of the computed assignment and adding more links does not change the overall network cost significantly.

\section{Related Work}
\label{sec:relatedwork}

Cloudlets~\cite{cloudlets} propose a small-scale, localized cloud installed at the edge of the network along with the centralized cloud and is based on virtualization technologies. Several other works have explored combining stable peer-resources as nano data centers, micro clouds, community clouds, etc., for compute/storage tasks~\cite{costcdn,serendipity,cloudletmobile,compoffloading}. 

Several researchers have proposed to bring part of the cloud closer to the edge of the network. Following the Fog cloud characteristics proposed by CISCO~\cite{cisco_fog_overview}, Bonomi et al.~\cite{bonomifog} and Yannuzzi et al.~\cite{fogcomp_iot} show that the fog is the appropriate platform for loosely coupled, computationally intensive IoT-based applications, such as connected vehicles and smart cities. Hong et al.~\cite{mobile_fog} provide a programming model and API for developing applications on the Fog cloud. On the other hand, unlike installing managed compute resources as Fog devices to process cloud applications, Lopez et al.~\cite{edge_ccr} propose a semi-centralized cloud architecture, Edge cloud, composing of volunteer-based, user-centric compute resources. Likewise, Ryden et al.~\cite{nebula_ccr} proposed a dispersed cloud, Nebula, which utilizes volunteer resources for running data-intensive tasks. The authors discuss the effectiveness of their approach by deploying Map-reduce jobs on available resources. 

Our work differs from all these approaches as unlike them, wherein a central entity schedules and processes several application tasks; Edge-Fog cloud proposes an entirely decentralized computing mechanism. Due to its unique nodular and layered architecture, the Edge-Fog cloud natively supports computations on distributed, semi-dependent data produced by IoT. 

\section{Conclusion} \label{sec:conclusion}

In this paper, we proposed the Edge-Fog cloud, a decentralized cloud model for handling computation-based, high volume and distributable data such as that generated by IoT. The model builds on the existing Edge and Fog cloud approaches and provides data resilience through a centralized data store. We also provided a novel task allocation mechanism for Edge-Fog cloud which significantly reduces the deployment time without sacrificing the associated cost when compared to related approaches. Further, we address several questions which might impact the real-world implementation of Edge-Fog cloud.

Future work in the area includes considering practical implementation and deployment issues of LPCF in a realistic Edge-Fog scenario.

%There is a vast scope of future work in Edge-Fog cloud. Currently we do not consider the effects on network cost while transferring the required data block from central repository to an Edge/Fog device. Further, providing a secure computation platform/VM at voluntary Edge devices is another important direction which needs to be explored. \textbf{Privacy}

\vskip -3mm

\section{Acknowledgement}

This research was funded by the joint EU FP7 Marie Curie Actions Cleansky Project, Contract No. 607584.
%

%
%\input{discussion}
%


%{
%\bibliographystyle{IEEEtran}
%\bibliography{IEEEabrv,new_references}
%}

\end{document}